\documentclass[10pt,conference]{IEEEtran}

\IEEEoverridecommandlockouts
\usepackage[utf8]{inputenc}
\usepackage{cite}
\usepackage{amsmath,amssymb,amsfonts}
\usepackage{algorithmic}
\usepackage{graphicx}
\usepackage{textcomp}
\usepackage[table]{xcolor}
\usepackage{url}
\usepackage{subfigure}
\usepackage{dblfloatfix}
\usepackage[all]{nowidow}

\usepackage{array,multirow}
\usepackage{float}
\usepackage{balance} 
\usepackage[inline,shortlabels]{enumitem}
 
 \newif\ifsubm

\submfalse

\usepackage{todonotes}

\def\BibTeX{{\rm B\kern-.05em{\sc i\kern-.025em b}\kern-.08em
    T\kern-.1667em\lower.7ex\hbox{E}\kern-.125emX}}

\makeatletter
\def\ps@IEEEtitlepagestyle{
	\def\@oddfoot{\mycopyrightnotice}
	\def\@evenfoot{}
}
\def\mycopyrightnotice{
	{\footnotesize
		\begin{minipage}{\textwidth}
			\centering
			\textcopyright~2019 IEEE. Personal use of this material is permitted.  Permission from IEEE must be obtained for all other uses, in any current or future media, including reprinting/republishing this material for advertising or promotional purposes, creating new collective works, for resale or redistribution to servers or lists, or reuse of any copyrighted component of this work in other works.
		\end{minipage}
	}
}
\usepackage[hidelinks]{hyperref}
\hypersetup{
	bookmarks=true,         
	unicode=false,          
	pdftoolbar=true,        
	pdfmenubar=true,        
	pdffitwindow=false,     
	pdftitle={Planning as Optimization: Dynamically Discovering Optimal Configurations for Runtime Situations},
	pdfauthor={Erik M. Fredericks, Ilias Gerostathopoulos, Christian Krupitzer, Thomas Vogel},
	pdfsubject={},
	pdfcreator={},
	pdfproducer={},
	pdfkeywords={planning, optimization, Bayesian optimization, evolutionary search, traffic routing model problem},
	pdfnewwindow=true,
}
    
\begin{document}

\title{Planning as Optimization: Dynamically Discovering Optimal Configurations for Runtime Situations\\[-.45em]
{\footnotesize \textsuperscript{*}Authors ordered alphabetically}\\[-.65em]
}

\author{
	\IEEEauthorblockN{
		Erik M. Fredericks\IEEEauthorrefmark{1},
		Ilias Gerostathopoulos\IEEEauthorrefmark{2},
		Christian Krupitzer\IEEEauthorrefmark{3},
		and
		Thomas Vogel\IEEEauthorrefmark{4}
	}
	\IEEEauthorblockA{\IEEEauthorrefmark{1}\textit{Oakland University}, Rochester, MI, USA. Email:~fredericks@oakland.edu}
	\IEEEauthorblockA{\IEEEauthorrefmark{2}\textit{Technical University of Munich}, Munich, Germany. Email:~gerostat@in.tum.de}
	\IEEEauthorblockA{\IEEEauthorrefmark{3}\textit{University of Würzburg}, Würzburg, Germany. Email:~christian.krupitzer@uni-wuerzburg.de}
	\IEEEauthorblockA{\IEEEauthorrefmark{4}\textit{Humboldt-Universität zu Berlin}, Berlin, Germany. Email:~thomas.vogel@cs.hu-berlin.de}
}

\maketitle

\begin{abstract}
The large number of possible configurations of modern software-based systems, combined with the large number of possible environmental situations of such systems, prohibits enumerating all adaptation options at design time and necessitates planning at run time to dynamically identify an appropriate configuration for a situation. While numerous planning techniques exist, they typically assume a detailed state-based model of the system and that the situations that warrant adaptations are known. Both of these assumptions can be violated in complex, real-world systems. As a result, adaptation planning must rely on simple models that capture what can be changed (input parameters) and observed in the system and environment (output and context parameters). We therefore propose planning as optimization: the use of optimization strategies to discover optimal system configurations at runtime for each distinct situation that is also dynamically identified at runtime. We apply our approach to CrowdNav, an open-source traffic routing system with the characteristics of a real-world system. We identify situations via clustering and conduct an empirical study that compares Bayesian optimization and two types of evolutionary optimization (\mbox{NSGA-II} and novelty search) in CrowdNav.
\end{abstract}

\begin{IEEEkeywords}
planning, optimization, Bayesian optimization, evolutionary search, traffic routing model problem
\end{IEEEkeywords}

\section{Introduction}

A self-adaptive system (SAS) continuously monitors itself and its environment to ensure that, for each environmental situation, a valid system configuration is applied that achieves optimal performance and behavior~\cite{Cheng2009a}. 
Using the monitored data, adaptation planning aims at identifying such a valid and optimal configuration.
For this purpose, a SAS integrates decision metrics based on rules or models with higher degrees of freedom~\cite{Krupitzer2015}.
However, both categories have their shortcomings. 
Rules can be inflexible if not accompanied with runtime learning~\cite{Rodrigues:2018} and they cannot cover all situations due to the state-space explosion related to the number of possible situations and configurations~\cite{Cheng2009, Cheng2009a, McKinley2004}. %
Models such as Discrete Time Markov Chains~\cite{Calinescu:2011} might cope with higher numbers of situations and configurations as they offer more adaptation freedom.
However, such models require detailed knowledge about the internal behavior of the system, the behavior of its environment, and the effects of adaptation actions to the system for all possible environmental situations.

In complex, real-world systems it is difficult to both identify the environmental situations that warrant adaptation and understand how changes to the system affect the performance and behavior in these situations.
For example, optimizing a router in a traffic system requires a detailed model of the behavior of the individual cars, the traffic events that may occur (e.g., increase of traffic demand), and how changes in routing affect the performance (e.g., the average trip time) and behavior (e.g., traffic jams). 
It is a challenge to obtain, maintain, and tailor such detailed models to each environmental situation.

An alternative approach is to model the system as a \textit{black-box} with input and output parameters and its environment as a set of context parameters. 
Optimizing the system then involves finding values for input parameters (i.e., \textit{configurations}) that optimize the system performance and behavior specified in terms of output parameters, and do so for each environmental situation specified in terms of context parameters. 
In other words, optimization is performed as a means of adaptation planning for each situation.  
To be effective as a method at runtime and applicable to real-world systems, optimization needs to cope with large numbers and ranges of input, output, and context parameters, and provide useful results in a timely manner. 
The last point is especially important for usage in scenarios where the optimization horizon is short. 

In this setting, different optimization techniques can be used if they can cope with problems that are
(i)~\textit{black-box}, i.e., the function relating the input to the output parameters is unknown,
(ii)~\textit{high-dimensional}, i.e., large number and ranges of input, output, and context parameters, 
and
(iii)~\textit{expensive}, i.e., they need to be solved in a minimal number of iterations since computing outputs may be costly in terms of time or other resources~\cite{Shan2010}. 
Examples of such optimization techniques include Bayesian and evolutionary approaches.

Another challenge is ``to support self-adaptation for complex types of uncertainties''~\cite[p.\,436]{Weyns2019}, i.e., when it is not possible to model \textit{a priori} the situations in which a SAS might reside at run time.
In theory, optimization at run time can solely focus on finding a system configuration that works well in any situation (e.g., to optimize the parameters of a web server without considering the fluctuations in demand). In practice, such a situation-agnostic optimization may lead to sub-optimal configurations. 
Therefore, a system should dynamically identify the \textit{distinct} situations it encounters at runtime, based on the effect of contextual parameters on the outputs, and optimize individually for each of them.
For example, a distinct situation in a traffic system may be a traffic jam or an accident.

Responding to these challenges, we propose an approach that we call \textit{planning as optimization}: the use of optimization strategies to discover optimal system configurations at runtime for each distinct situation that is dynamically identified at runtime.
Our approach draws inspiration from online optimization and learning \cite{porter_losing_2016,Pilgerstorfer+Evangelos:2017,jiang_pytheas:_2017} and
tackles complex systems that are modeled as black-box, high-dimensional, and computationally expensive optimization problems.

In particular, we make the following contributions:
\begin{itemize}
    \item We show the feasibility of planning as optimization by dynamically identifying distinct situations at runtime via clustering and by discovering optimal configurations via different optimization techniques.
    \item We compare the solution quality, convergence, and overhead of three optimization techniques in an empirical study using CrowdNav~\cite{Schmid2017}, an open-source self-adaptation exemplar of a traffic system that corresponds to a black-box, high-dimensional, and expensive optimization problem. 
\end{itemize}

As optimization techniques we select Bayesian optimization and two evolutionary optimization techniques (NSGA-II and novelty search) as they can cope with such a problem.
Since in general no technique is superior to any other technique for any given optimization problem (\textit{cf.}\,no ``free lunch" theorems for optimization~\cite{Wolpert:1997}), we performed the comparison between the three techniques to identify which performs best in CrowdNav. %
This is also a first step towards our vision of dynamically identifying and using the best optimizer at runtime.

The rest of this paper is organized as follows: 
In Section~\ref{sec:relWork}, we illustrate the research gap with a literature review.
Section~\ref{sec:scenario} describes our motivating scenario based on CrowdNav.
We present our planning as optimization approach involving situation detection and optimization with three optimizers 
in Section~\ref{sec:approach}.
We evaluate our approach in an empirical study in Section~\ref{sec:eval} and discuss remaining research challenges in Section~\ref{sec:implications}.
Lastly, we summarize our findings in Section~\ref{sec:conclusion}.

\section{Using Optimization in Self-adaptive Systems}
\label{sec:relWork}

Many approaches apply optimization in SASs for adaptation planning by generating new system configurations or adaptation plans.
We analyzed these techniques based on approaches published during the last ten years in conferences and journals related to SASs.\footnote{We considered ACM TAAS, ICAC, SASO, SEAMS, and FSE. Due to space constraints we did not include the references in this paper but published them here: \url{https://doi.org/10.5281/zenodo.2584266}.}
As listed in Table~\ref{tab:techniques}, we identified the use of 29 different techniques in 51 publications. This list shows that a large set of techniques from different classes such as 
probabilistic,
combinatorial,
evolutionary,
stochastic,
mathematical,
and
meta-heuristic
optimization are applied in SASs.
However, there is less information on how the techniques compare to each other in terms of assumptions, overhead, and quality of the achieved solutions.

\begin{table}[t!]
	\caption{List of Optimization techniques used in SASs.}
	\label{tab:techniques}
	\vspace{-2mm}
	\centering
	\begin{tabular}{|p{0.95\columnwidth}|}
		\hline
		\textbf{Probabilistic Optimization Techniques}\\
		Bayesian Networks, %
		Bayesian Optimization, %
		Simulated Annealing\\%
		\hline
		\textbf{Combinatorial Optimization Techniques} \\
		Cross-entropy Method for Combinatorial Optimization, %
		Decentralized Combinatorial Optimization\\%
		\hline
		\textbf{Evolutionary Optimization Techniques} \\
		Evolutionary Algorithm, %
		Genetic Algorithm, %
		Genetic Programming, %
		Learning Classifier System, %
		NSGA-II, %
		SPEA2\\%
		\hline
		\textbf{Stochastic Optimization Techniques} \\
		Greedy Algorithm, %
		Markov Decision Process, %
		Stochastic Approximation, %
		Stochastic Programming, %
		Variable Neighbourhood Search\\%
		\hline
		\textbf{Mathematical Optimization Techniques}\\
		Binary Programming, %
	    Integer Programming, %
	    Linear Programming, %
	    Sequential Quadratic Programming, %
	    Convex Optimization Solver, %
	    Pattern Search Algorithm\\%
		\hline
		\textbf{Meta-Heuristic Optimization Techniques}\\
		Heuristic Algorithm, %
		Tabu Search\\%
		\hline
		\textbf{Other Optimization Techniques} \\
		Canonical Correlation, %
		Weighted Sum Model, %
		Reinforcement Learning, %
		Distributed Constraint Optimization, %
		Gradient Descent\\%
		\hline
	\end{tabular}
	\vspace{-1em}
\end{table}

\begin{table*}[!t]
\caption{Input, Output, and Context Parameters of CrowdNav.}
\vspace{-2mm}
\resizebox{\textwidth}{!}{%
\begin{tabular}{llll}
\multicolumn{4}{c}{\cellcolor{gray!25}\textit{\textbf{Input Parameters}}}                                                                                                     \\ \hline
\textbf{Name}            & \textbf{Type} & \textbf{Range}       & \textbf{}                                                                                \\ 
route randomization      & float         & {[}0-0.3{]}          & Controls the random noise introduced to avoid giving the same routes                     \\ 
exploration percentage   & float         & {[}0-0.3{]}          & Controls the ratio of smart cars used as explorers                                       \\ 
static info weight       & float         & {[}1-2.5{]}          & Controls the importance of static information (i.e., max speed, street length) on routing \\ 
dynamic info weight      & float         & {[}1-2.5{]}          & Controls the importance of dynamic information (i.e., observed traffic) on routing         \\ 
exploration weight       & integer       & {[}5-20{]}           & Controls the degree of exploration of the explorers                                      \\ 
data freshness threshold & integer       & {[}100-700{]}        & Threshold for considering traffic-related data as stale and disregard them               \\ 
re-routing frequency     & integer       & {[}10-70{]}          & Controls how often the router should be invoked to re-route a car                   \\ \hline
\multicolumn{4}{c}{\cellcolor{gray!25}\textit{\textbf{Output Parameters}}}                                                                                                              \\ \hline
\textbf{Name}            & \textbf{Type} & \textbf{Constraints} & \textbf{Description}                                                                     \\
trip overhead            & float         & \textgreater 1       & The actual trip time divided by the theoretical trip time if a car travels at max speed   \\
routing cost             & integer       & \textgreater 0       & The time needed by the router to re-route all cars \\  \hline
\multicolumn{4}{c}{\cellcolor{gray!25}\textit{\textbf{Context Parameter}}}                                                                                                              \\ \hline
\textbf{Name}            & \textbf{Type} & \textbf{Constraints} & \textbf{Description}                                                                     \\
number of cars          & integer       & \textgreater 0       & The total number of cars that are in the city and use the smart configurable router \\ 
\hline           
\end{tabular}%
}
\label{tab:crowdnav}
\vspace{-1em}
\end{table*}

Our overall approach of finding optimal configurations in different situations bears similarities with the work by Porter~\textit{et~al.} on learning optimal 
system configurations in emergent software systems~\cite{porter_losing_2016}. 
This work focuses on providing a general framework for online learning that interleaves exhaustive search of configurations with determination of environmental situations. 
In contrast, our work focuses on optimization problems of higher dimensionality that require more sophisticated search techniques.
Similar to our work, Kinneer~\textit{et al.}~\cite{Kinneer2018} introduced an approach for adaptation planning that re-uses the knowledge of existing plans for optimization via genetic algorithms.
In contrast to our approach, the authors rely on tactics that can change the system by integrating explicit knowledge of the system, whereas we target black-box systems where such knowledge is not available.
Further, different authors did studies to compare the performance of optimization techniques.
Bischl~\textit{et~al.} compare mlrMBO, a flexible toolbox for black-box optimization 
with Bayesian optimization~\cite{Bischl2018} against other optimization techniques, including NSGA-II.
However, they used theoretical problems for their comparison whereas we apply the optimization techniques in a (simulated) traffic system that represents a real-world problem. %
In another study, Moreno~\textit{et~al.}~\cite{Moreno2017_study} compare Markov Decision and Analytic Hierarchy Processes for adaptation planning in \textit{RUBiS}.
However, both techniques are not applicable to black-box systems that we are targeting.

\section{Motivating Scenario}
\label{sec:scenario}

Many systems have been modeled as SASs in domains such as cyber-physical~\cite{fredericks.2018.icdcs,bencomo.2013} and intelligent traffic systems~\cite{Tomforde2008,Wuttke2012}.
In this paper, we focus on intelligent traffic systems and
perform an empirical study on the \mbox{CrowdNav}\footnote{CrowdNav: \url{https://github.com/iliasger/CrowdNav}} exemplar~\cite{Schmid2017},  a SAS that performs smart routing for city-wide traffic management.
CrowdNav combines the SUMO traffic simulator\footnote{Simulation of Urban MObility:  \url{http://sumo.dlr.de/index.html}} with a custom-built module for routing. 
In CrowdNav, cars continuously drive in the city of Eichst{\"a}dt from a randomly selected destination to another with routes provided by the routing module.
The module is used by all cars and comprises seven configurable numeric input parameters detailed in Table~\ref{tab:crowdnav}.
Moreover, CrowdNav provides two outputs, \textit{trip overhead} and \textit{routing cost}, 
and features one context parameter that can be observed but not controlled: the \textit{number of cars}~(see~Table~\ref{tab:crowdnav}). 

Selecting an optimal configuration in CrowdNav entails providing a value for each input parameter (\textit{route randomization}, \textit{exploration percentage}, etc.) with the goal of minimizing \textit{trip overhead} and \textit{routing cost} for each situation determined by the \textit{number of cars}.  
Note that a configuration impacts each output parameter differently, leading to a multi-objective optimization problem with competing concerns.

\begin{figure}[t]
	\centering
	\includegraphics[width=0.44\textwidth]{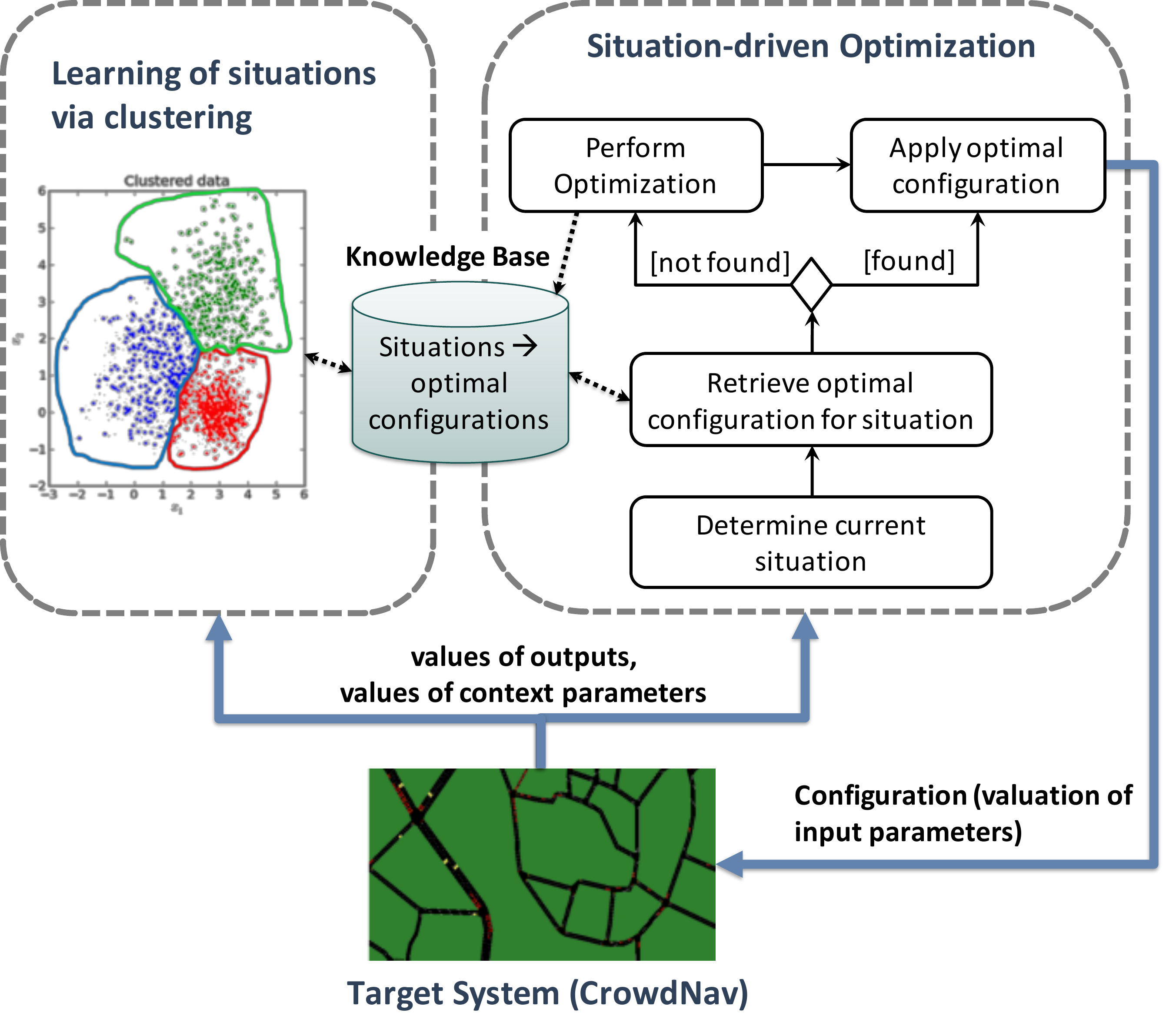}
	\vspace{-1em}
	\caption{Overview of our approach.}
	\label{fig:approach}
	\vspace{-1em}
\end{figure}

Viewed as a numeric optimization problem, CrowdNav optimization has some distinct properties. 
First, it is a black-box optimization problem since there is no known model/function that relates inputs to outputs. 
As a result, to evaluate a configuration, it needs to be applied to CrowdNav and its effects need to be measured on the outputs.
Second, the number and range of input and context parameters in CrowdNav also creates a large configuration space, resulting in a high-dimensional problem.
Given that a float is represented in Python by default with 15 decimal points and assuming that no discretization is performed, the input configuration space of CrowdNav is $0.3\cdot10^{15}\cdot0.3\cdot10^{15}\cdot1.5\cdot10^{15}\cdot1.5\cdot10^{15}\cdot15\cdot600\cdot60\approx100,000\cdot10^{60}$.
Third, the outputs of CrowdNav exhibit high variance (\textit{noisy} outputs).
To compare different configurations as to their effects on the outputs,  multiple samples of the outputs are required to ensure that statistical measures are robust to noise and outliers. 
We found that 5000 samples are sufficient to characterize a situation and to evaluate a configuration.
However, collecting multiple samples from a running system increases the time needed to evaluate a configuration, which makes CrowdNav an expensive optimization problem.

Finally, optimization of CrowdNav must consider different situations that depend on the value of its context parameter, \textit{number of cars}, as it is unlikely that a single optimal configuration exists for different values (e.g., 300 vs. 800 cars).

\section{Approach}
\label{sec:approach}

Our approach aims at optimizing systems that are modeled as black-box, high-dimensional, and computationally expensive optimization problems in different runtime situations.
Therefore, there is a need for both learning the distinct situations the system might be in, and optimizing the system for each such situation. 
Our approach comprises two modes, each dealing with one of the above-mentioned needs (Figure~\ref{fig:approach}).

\noindent\textbf{Mode \#1: Learning of situations via clustering.}
In this mode, the outputs of the target system and the context parameters are monitored. The goal is to determine the valuations of context parameters that  
can be grouped together in a distinct \textit{situation} in terms of the outputs. 
Once these valuations are learned, the system can detect its current situation by only monitoring its context parameters. %

To learn a situation, we assume that for each context parameter, a discrete number of ranges for its values is provided.
For instance, the designers of CrowdNav may specify that the \textit{number of cars} -- a context parameter in CrowdNav -- belongs to one of the ranges [0,100], [100, 200], [200, 300], [300, $\infty$).
All possible states that the system can reside in is the Cartesian product of the ranges of all of its context parameters. 
While operating, the system can traverse from one state to another. 
By observing the values of its output parameters for each state, our approach essentially groups together states into situations by learning which states are similar to each other with respect to their impact on the output parameters. 
The clustering of the output data to learn situations is detailed in Section~\ref{sec:analyzer}.

\noindent\textbf{Mode \#2: Situation-driven optimization.}
In the second mode, the context parameters of the target system are monitored and the clustering model learned in \textit{Mode \#1} is used to determine the current situation.  
If the current situation is different from the previous one, the \texttt{Knowledge Base} (\textit{cf.} Figure~\ref{fig:approach}) is queried for an optimal configuration. 
If such a configuration is already known, it is applied to the target system.  
If not, an optimization process starts with the goal of identifying an optimal configuration for the current situation, applying the optimized configuration to the target system, and saving it to the \texttt{Knowledge Base} for future use.

Situation-driven optimization relies on the learning of situations via clustering to optimize for distinct situations.
If clustering is not performed or returns just a single situation (indicating that the context does not influence the outputs), optimization can still be performed for this general case. 

We assume that an optimization process is not interrupted once started so that it finishes before the current situation changes. 
If this assumption does not hold, the system needs to be equipped with a mechanism of saving the currently best configuration for a situation to the \texttt{Knowledge Base} and continuing the optimization process when it is next detected.
Such an incremental optimization is a topic of future work.

The optimization process of our approach can be guided by different optimization techniques. 
So far, we considered three state-of-the-art techniques in optimizing CrowdNav  (\textit{cf.}~Section~\ref{sec:optimizationTechniques}). 
Generally, the choice of the technique highly depends on the target system and in particular in the response surface of its outputs. 
We therefore provide a basis for comparison between multiple optimization techniques (\textit{cf.}~Section~\ref{sec:eval}). 

We next discuss how distinct situations are learned via clustering as well as the optimization techniques we~use.

\subsection{Clustering-based Situation Learning}
\label{sec:analyzer}

To group individual context states to situations, our approach continuously observes both the valuation of context parameters and the corresponding system outputs. 
For each context state (the number of context states is the Cartesian product of all the possible ranges of context parameters), a number of observations of system outputs are collected. 
This collection results in a dataset for each context state, with features computed for each dataset. 
Possible features include well-known statistical measures of central tendency and dispersion such as arithmetic mean, median, variance, and standard deviation. 
Our approach assumes that such features are provided for each system output (however, we can always use generic features, such as the statistical measures mentioned above). 
The features for each context state are then fed to a clustering algorithm -- we use \textit{k-means}~\cite{na_research_2010} -- that determines the datasets that are most similar and should form a cluster. 
In particular, given a number of clusters \textit{k}, \textit{k-means} iteratively tries to find a centroid for each cluster so that the sum of the squared Euclidean distances between an observation assigned to the cluster and the cluster's centroid is minimized.  
The context states that correspond to the datasets belonging to the same cluster are then grouped together in a~situation. 

For illustration with CrowsNav, consider that the arithmetic mean and median of \textit{trip overheads} are used as clustering features. Moreover, the only context variable is the \textit{number of cars} that can be in one of the four ranges [0,100], [100, 200], [200, 300], [300, $\infty$) corresponding to four system states. Our clustering approach will compute the mean and median of a large number of samples (5000) of trip overhead for each state. 
If \textit{k-means} groups the first three states into a single cluster, then there is not enough difference between having 100 or 300 cars, corresponding to a single situation (e.g., ``low traffic"). 

A problem with using \textit{k-means} or any other clustering algorithm at runtime is that these algorithms typically expect the \textit{number of clusters} to be provided by the user. 
Instead, we assume that a list of candidate numbers of clusters is provided, from which the optimal number of clusters is automatically determined at runtime based on the available data. 
To determine the optimal number of clusters, we follow the Silhouette method~\cite{ROUSSEEUW198753}, a well-accepted method for measuring clustering validity~\cite{starczewski_performance_2015}. 
For each number of clusters, we perform clustering via \textit{k-means} and then compute the average Silhouette coefficient. 
In particular, the Silhouette coefficient $sc$ for a datum is calculated by Equation~\ref{eq:silhouette}:
\begin{equation}
    sc = \frac{b - a}{max(a, b)}
    \label{eq:silhouette}
\end{equation}
where $a$ is the average Euclidean distance between the datum and other data in its cluster and $b$ is the average Euclidean distance between the datum and other data in the next nearest cluster. 
A Silhouette coefficient takes values within [-1,1] with values close to 1 indicating a good match of the datum to the cluster. 
The average Silhouette coefficient is calculated by considering all data points and provides a measure of how well the data are assigned to clusters~\cite{starczewski_performance_2015}.
In our approach, we select the number of clusters that yields the highest average Silhouette coefficient.

\subsection{Optimization Techniques}
\label{sec:optimizationTechniques}

Since each optimization technique presented in Section~\ref{sec:relWork} has its own strengths and limitations, they cannot all be applied to all optimization problems. %
For a black-box problem/system, there is generally no model describing its function/behavior. This limitation rules out all techniques that require such a model, such as linear programming.
Moreover, a problem with high dimensionality poses challenges for techniques that do not scale and a multi-objective problem prevents the use of single-objective techniques.

For this paper, we consider probabilistic and evolutionary techniques to optimize CrowdNav, as it is a black-box system comprising high-dimensionality and multi-objective characteristics.  These techniques have been shown to cope with such criteria in SASs (\textit{cf.}~Section~\ref{sec:relWork}) in related optimization problems.
Particularly, as a probabilistic technique we use \textit{Bayesian Optimization} that has been previously applied to CrowdNav for a single-objective problem~\cite{Gerostathopoulos2018} and as evolutionary techniques we use the widespread \mbox{\textit{NSGA-II}} algorithm and \textit{novelty search}. 
These three techniques all rely on some form of \textit{fitness functions} to evaluate a configuration of the CrowdNav router in terms of its objectives: \textit{trip overhead} and \textit{routing cost}. Having no model of CrowdNav, we have no means to directly calculate the fitness of a router configuration. Instead, a configuration is applied to CrowdNav and its effects on the \textit{trip overhead} and \textit{router cost} are measured to obtain the fitness. This corresponds to an \textit{online experiment} of applying and evaluating a router configuration in the running \mbox{CrowdNav}.
Next, we briefly describe the three optimization~techniques we employed.

\subsubsection{Bayesian Optimization}

Bayesian or sequential model-based optimization is an approach to global optimization that can be used for efficiently optimizing expensive black-box functions~\cite{shahriari_taking_2016}. 
By expensive it is meant that a single evaluation of the function is costly in terms of time or resources. 
Bayesian optimization can be used for optimizing a single objective (already demonstrated on CrowdNav~\cite{Gerostathopoulos2018}) as well as a multi-objective problem, which is the focus of this paper.  

In short, Bayesian optimization works as follows. 
Given a number of execution steps (\textit{budget}), at each step, the process fits a regression model to the selected inputs and obtained outputs, then uses the model to propose a promising set of inputs to try next by optimizing an \textit{acquisition function}. 
A common approach for the regression model is to use Gaussian processes. Such processes can capture the uncertainty in the measurements and deal with noisy functions. They have been applied to CrowdNav whose outputs have high variance~\cite{Gerostathopoulos2018}. 

Being a topic of active research, many different flavors of Bayesian multi-objective optimization have been proposed~\cite{gaspar-cunha_model-based_2015}.
They differ mainly on whether they use a single (i.e., scalarized objectives) or separate regression models.
We have chosen a variant called \textit{$\mathcal{S}$-Metric-Selection-based Efficient Global Optimization} (SMS-EGO)~\cite{Ponweiser2008}.
In this algorithm, separate regression models for each objective are fitted and the proposed point to evaluate next is selected based on the estimated contribution to the hypervolume indicator.

\subsubsection{NSGA-II}
The Non-dominated Sorting Genetic Algorithm II (NSGA-II) is a multi-objective evolutionary algorithm that searches for pareto-optimal solutions to an optimization problem~\cite{Deb2002}. During the search, NSGA-II evolves a population of candidate solutions using crossover, mutation, and selection operators inspired by evolution and natural selection in biology. The goal is to find solutions that are optimal with respect to the search objectives. For this purpose, a fitness function is used that evaluates how well a solution satisfies the objectives. The resulting fitness of a solution determines the selection of this solution to the next generation for further evolution steps. Having multiple objectives, the result of the search is a pareto frontier, a set of solutions with the best trade-offs between the objectives that could be found.

Additionally, NSGA-II promotes the diversity of solutions, which supports exploring the search and objective spaces. In contrast to the original NSGA, it introduces elitism that avoids losing good solutions during the search and improves the performance of the non-dominated sorting. These aspects make NSGA-II a popular technique that is widely used in optimization as well as search-based software engineering~\cite{Harman:2012}.

For CrowdNav, a configuration of the router (i.e., a candidate solution) is encoded as a chromosome or a vector of components with one component for each input parameter. While mutation randomly modifies the value of one input parameter taking the defined range of this parameter into account, crossover recombines two configurations to obtain a new configuration. To evaluate the fitness of a single configuration in terms of \textit{trip overhead} and \textit{routing cost}, an online experiment is performed in the running CrowdNav.

\subsubsection{Novelty Search}  

Novelty search provides an alternative method for evolutionary optimization by searching the solution space for uniquely optimal, rather than solely optimal, solutions.  In contrast to more common evolutionary processes (e.g., genetic algorithms), novelty search relies on a measure of distance between genomes as a point of optimization while considering the validity and/or optimality of the genomes, commonly called the \textit{novelty metric}.  The intent of the novelty function is to avoid the issue in which an evolutionary process becomes ``stuck'' in a local optima and instead explores the solution space for a globally-optimal solution~\cite{Lehman2008}.

Generally, the novelty metric is calculated from a combination of the combined pair-wise distances between all generated solutions (e.g., Manhattan distance between instantiated genome parameters) and a measure of performance for each solution (e.g., the fitness of solutions).  These values are combined via a linear-weighted sum into the overall novelty metric that quantifies the diversity, combined with the optimality, of each solution.  For this instantiation of novelty search, we use the same approach for defining genomes as does our implementation of NSGA-II.  %

Novelty search also differs from common evolutionary approaches in that it maintains a \textit{novelty archive} of the most diverse solutions.  This archive is populated each generation by ranking all genomes in the population, as well as the contents of the archive, and selecting the $k$ most diverse solutions (i.e., the solutions with the highest novelty score).  Upon completion, the novelty archive will contain the $k$ most diverse solutions discovered  throughout the entirety of the search.  For this paper, we set $k$ to retain the top $20\%$ of all evaluated solutions.   %

\section{Evaluation}
\label{sec:eval}

In this section, we evaluate the two modes of our approach, learning of situations and situation-driven optimization.

\subsection{Experimental Setup}

Using CrowdNav as a managed system, we investigate the learning of situations by clustering and how different optimization techniques for adaptation planning perform in identifying optimal configurations for these situations. The optimization techniques are Bayesian Optimization, \mbox{NSGA-II}, and novelty search.  
To connect the optimization techniques to CrowdNav, we use \textit{RTX},\footnote{RTX is available open source: \url{https://github.com/iliasger/RTX/}} a framework that supports online experiments with CrowdNav. Thus, the three optimization techniques have been implemented in RTX.\footnote{\url{https://github.com/iliasger/RTX/tree/saso19}}

We ran the experiments comparing the different optimization techniques on identical virtual machines.\footnote{The virtual machines run Ubuntu 18.04, 16 Intel Haswell vCPUs, and 14.4gb of memory, and they host CrowdNav in version \url{https://github.com/iliasger/CrowdNav/tree/saso19} with SUMO in version 0.32.0.}
For each experiment, we use the Wilcoxon-Mann-Whitney U-test ($p < 0.05$) to determine statistical differences between datasets, as we assume no normality of data.  To establish statistical significance, we performed $30$ replicates for each experiment~\cite{Arcuri:2014}, where the replicate served as the seed value to RTX/CrowdNav.

\subsection{Experiments on Learning of Situations via Clustering}
\label{sec:evalClustering}

\begin{figure}[t]
	\centering
	\includegraphics[width=0.35\textwidth,page=1]{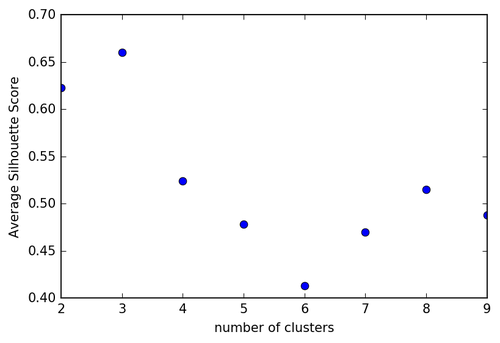}	
	\vspace{-0.2in}	
	\caption{Numbers of clusters and their score (the higher, the better).}
	\label{fig:avg_sil}	
	\vspace{-0.25in}
\end{figure}

To show the feasibility of dynamically identifying distinct situations at runtime via clustering, we performed an empirical study with CrowdNav, in which
we experimented with different values of the context parameter \textit{number of cars}.
Assuming that the ranges provided to the method for this parameter are [100-150], (150-200], ..., (750-800], we performed 14 experiments where each experiment had 150, 200, 250, ..., 750, 800 number of cars respectively, assuming that the highest value of a range is representative of all other values in the range. 
For each experiment, we collected 5000 samples of \textit{trip overhead} and calculated the average, median, 75th percentile, 90th percentile, standard deviation, and variance of each dataset. 
Then, we invoked \textit{k-means} with all the features for each dataset required to produce clusters with numbers in the range [2,9]. 
For each of the 8 produced clusterings, we computed the average Silhouette score. 
The results are depicted in Figure~\ref{fig:avg_sil}.
As can be seen, the clustering with three clusters had the highest score and thus was selected by our approach. 
Such a clustering groups the values of the context parameter into the following three groups: 
\begin{itemize}
    \item Cluster/Situation \#1: Car counts between 101 and 500.
    \item Cluster/Situation \#2: Car counts between 501 and 700.
    \item Cluster/Situation \#3: Car counts between 701 and 800. 
\end{itemize}

These three clusters correspond to ``low traffic,'' ``medium traffic,'' and ``high traffic,'' respectively. 
We note here that an extension of these experiments to also consider the features of \textit{routing cost} is straightforward. Next, we explain how our approach tries to optimize both CrowdNav objectives in each of the above situations by setting the number of cars to its highest possible value for each situation (i.e., 500, 700, and 800 cars, respectively).

\subsection{Experiments on Situation-driven Optimization}

To show the feasibility of discovering optimal configurations via different optimization techniques and evaluate these techniques in terms of solution quality, convergence, and overhead, we performed the following empirical study with CrowdNav.
For each of the three distinct situations of 500, 700, and 800 cars (\textit{cf.} previous section), we compare the three optimization techniques -- Bayesian optimization (BOGP), NSGA-II, and novelty search -- with each other and with random search as a baseline (\textit{cf.}\,\cite{Arcuri:2014}).
We configured each technique to perform 100 fitness evaluations, that is, 100 router configurations were generated, applied, and evaluated on CrowdNav during the optimization process.
For BOGP, this corresponds to a budget of 100.
For NSGA-II and novelty search, a population of size 10 is evolved over 10 generations with an offspring size of 10. Thus, each member of the population is adapted by mutation or crossover in each generation, resulting in a total of 100 candidate solutions evaluated throughout the search. The crossover and mutation rates are set to 0.7 and 0.3, respectively. Moreover, for novelty search the novelty archive size is set to 20\%. %
To provide a fair comparison, random search evaluated 100 randomly-generated candidate solutions.

Given that an evaluation of a single configuration takes approximately six minutes in our setting, if ten configurations are evaluated in parallel (e.g., for NSGA-II evaluating all ten configurations of one generation concurrently), then 100 evaluations can be performed within a given optimization horizon of 60 mins.
In our experiments, we did not employ any parallelization to better track the overall process. We now discuss the results from 30 replicates of running each optimization technique for each of the three situations.

\subsubsection{Solution Quality}

\newcommand{\subfigwidth}{5.0cm}
\newcommand{\save}{\vspace{-1mm}}
\begin{figure*}[htb!]%
\begin{center}
  \subfigure[Trip Overhead for 500 cars.]{%
  \label{sf:overhead-500}%
  \includegraphics[width=\subfigwidth]{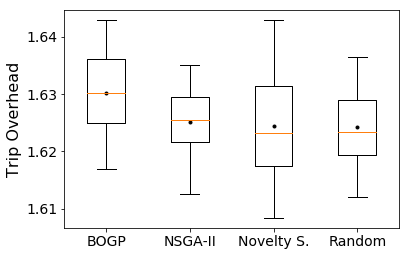}}%
  \qquad
  \subfigure[Trip Overhead for 700 cars.]{%
  \label{sf:overhead-700}%
  \includegraphics[width=\subfigwidth]{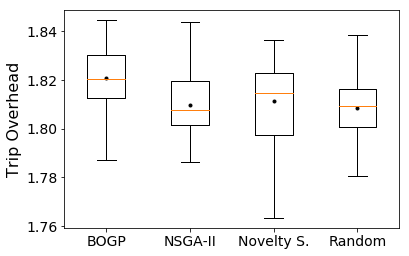}}%
  \qquad\save
  \subfigure[Trip Overhead for 800 cars.]{%
  \label{sf:overhead-800}%
  \includegraphics[width=\subfigwidth]{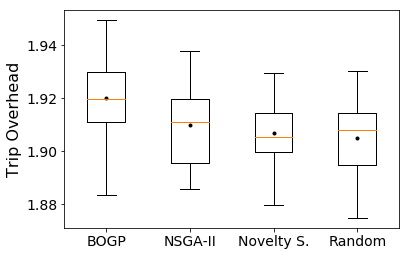}}%
  \qquad\save
  \subfigure[Routing Cost for 500 cars.]{%
  \label{sf:cost-500}%
  \includegraphics[width=\subfigwidth]{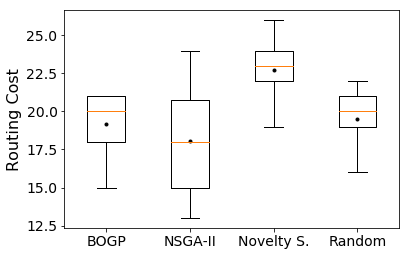}}%
  \qquad\save
  \subfigure[Routing Cost  for 700 cars.]{%
  \label{sf:cost-700}%
  \includegraphics[width=\subfigwidth]{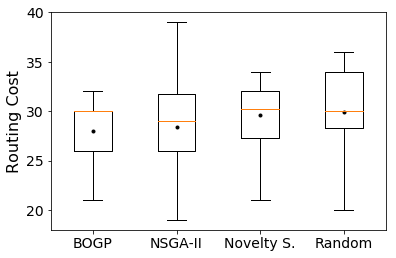}}%
  \qquad\save
  \subfigure[Routing Cost  for 800 cars.]{%
  \label{sf:cost-800}%
  \includegraphics[width=\subfigwidth]{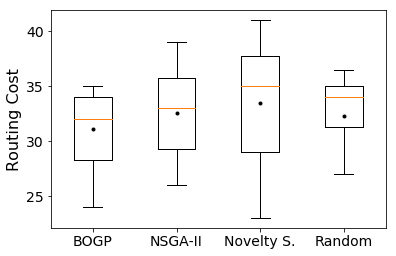}}%
  \qquad\save
  \subfigure[Hypervolume for 500 cars.]{%
  \label{sf:hv-500}%
  \includegraphics[width=\subfigwidth]{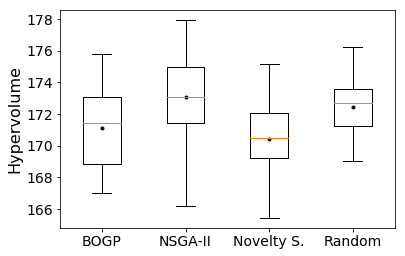}}%
  \qquad\save
  \subfigure[Hypervolume for 700 cars.]{%
  \label{sf:hv-700}%
  \includegraphics[width=\subfigwidth]{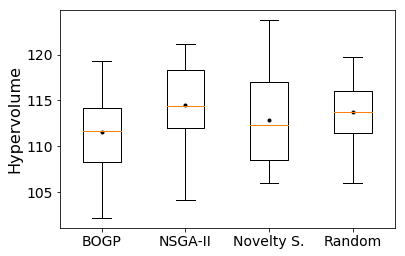}}%
  \qquad\save
  \subfigure[Hypervolume for 800 cars.]{%
  \label{sf:hv-800}%
  \includegraphics[width=\subfigwidth]{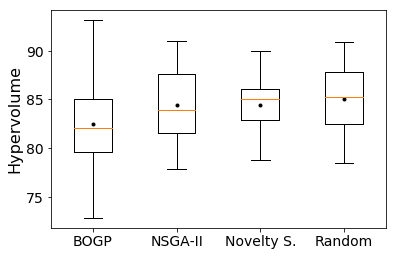}}%
  \qquad\save
  \subfigure[Hypervolume Evolution for 500 cars.]{%
  \label{sf:hve-500}%
  \includegraphics[width=\subfigwidth]{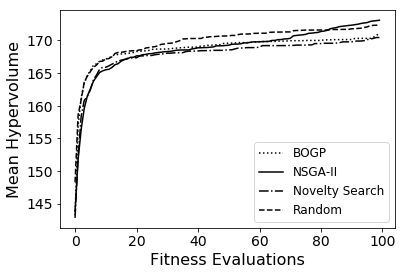}}%
  \qquad\save
  \subfigure[Hypervolume Evolution for 700 cars.]{%
  \label{sf:hve-700}%
  \includegraphics[width=\subfigwidth]{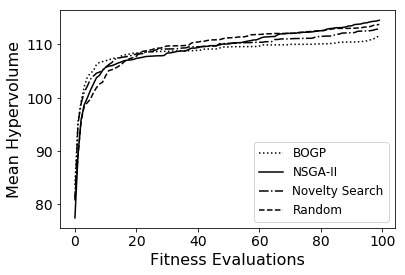}}%
  \qquad\save
  \subfigure[Hypervolume Evolution for 800 cars.]{%
  \label{sf:hve-800}%
  \includegraphics[width=\subfigwidth]{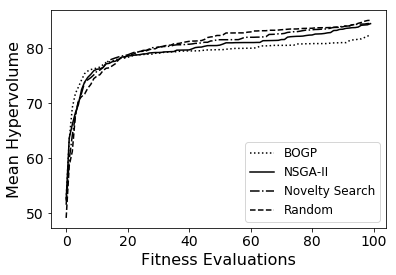}}%
  

  \caption{Trip Overheads \& Routing Costs (the lower, the better), and Hypervolumes (the higher, the better) for 500, 700, and 800 cars. Dots represent averages.}
  \label{f:quality}
  \vspace{-1.5em}
\end{center}
\end{figure*}

To evaluate how well an optimization technique performs, we consider the quality of the pareto-optimal router configurations found by the technique. The quality of a configuration is measured by how well the objectives, \textit{trip overhead} and \textit{routing cost}, are minimized.

Considering each objective individually, we selected the minimum value of \textit{trip overhead} and \textit{routing cost} achieved by the technique's pareto-optimal configurations. The corresponding trip overheads and routing costs over the 30 replicates are plotted in Figures~\ref{sf:overhead-500}-\subref{sf:cost-800} for each of the three distinct situations of 500, 700, and 800 cars. 
The averages and medians of trip overhead and routing cost are also listed in Table~\ref{tab:results} showing that the average and median values do not differ much. Thus, we just use the average values in the following.

\begin{table}[h]
\caption{Trip Overheads, Routing Costs, and Hypervolumes.}
\vspace{-2mm}
\resizebox{\columnwidth}{!}{%
\begin{tabular}{llrrrrrr}
& \textbf{Technique} & \multicolumn{2}{c}{\textbf{Trip Overhead}} & \multicolumn{2}{c}{\textbf{Routing Cost}} & \multicolumn{2}{c}{\textbf{Hypervolume}} \\
& & Average & Median & Average & Median & Average & Median \\ \hline
\parbox[t]{2mm}{\multirow{4}{*}{\rotatebox[origin=c]{90}{\textbf{500 cars}}}} & 
  BOGP            & 1.6302 & 1.6301 & 19.20 & 20.00 & 171.1091 & 171.4032 \\ 
& NSGA-II         & 1.6250 & 1.6254 & \cellcolor{gray!25}18.03 & \cellcolor{gray!25}18.00 & \cellcolor{gray!25}173.0898 & \cellcolor{gray!25}173.0635 \\
& Novelty Search  & 1.6244 & \cellcolor{gray!25}1.6232 & 22.73 & 23.00 & 170.4439 & 170.4801 \\
& Random          & \cellcolor{gray!25}1.6242 & 1.6234 & 19.53 & 20.00 & 172.4161 & 172.6764 \\ \hline 
\parbox[t]{2mm}{\multirow{4}{*}{\rotatebox[origin=c]{90}{\textbf{700 cars}}}} & 
  BOGP            &	1.8206 & 1.8202	& \cellcolor{gray!25}28.03 & 30.00	& 111.6234 & 111.6794 \\ 
& NSGA-II         &	1.8095 & \cellcolor{gray!25}1.8078 & 28.45 & \cellcolor{gray!25}29.00 & \cellcolor{gray!25}114.5563 & \cellcolor{gray!25}114.4319 \\
& Novelty Search  & 1.8112 & 1.8148 & 29.62 & 30.25 & 112.9054 & 112.2987 \\
& Random          &	\cellcolor{gray!25}1.8086 & 1.8093 & 29.90	& 30.00 & 113.7743 & 113.7692 \\ \hline 
\parbox[t]{2mm}{\multirow{4}{*}{\rotatebox[origin=c]{90}{\textbf{800 cars}}}} & 
  BOGP            &	1.9201 & 1.9195 & \cellcolor{gray!25}31.07 & \cellcolor{gray!25}32.00 & 82.4267 & 82.0824 \\
& NSGA-II         & 1.9098 & 1.9108 & 32.53 & 33.00	& 84.3859 & 83.9632 \\ 
& Novelty Search  & 1.9068 & \cellcolor{gray!25}1.9052 & 33.50 & 35.00 & 84.4170 &	85.0677 \\
& Random          & \cellcolor{gray!25}1.9049 & 1.9079 & 32.32 & 34.00	& \cellcolor{gray!25}85.0024 &	\cellcolor{gray!25}85.2451 \\ \hline
\end{tabular}%
}
\label{tab:results}
\vspace{-1.5em}
\end{table}

For all situations, random search, NSGA-II, and novelty search found configurations that achieve similar average trip overheads of around 1.62 (500 cars), 1.81 (700 cars), and 1.91 (800 cars) while BOGP is slightly worse with values of 1.63, 1.82, and 1.92, respectively.
Concerning the average routing cost in the situation with 500 cars, NSGA-II obtained the best results (18.03), followed by BOGP (19.20), Random (19.53), and novelty search (22.73). 
In contrast, for 700 cars, BOGP discovered the best results (28.03) followed by NSGA-II (28.45), novelty search (29.62), and random search (29.90). 
Likewise for 800 cars, BOGP found again the best configurations (31.07) followed by random search (32.32), NSGA-II (32.53), and novelty search (33.50).

Using the Wilcoxon-Mann-Whitney U-test ($p<0.05$) and concerning the \textit{trip overhead}, we observe a statistically significant difference
for 500 
and 700 cars between
\mbox{NSGA-II} and BOGP,
novelty search and BOGP,
as well as random search and BOGP;
and for 800 cars between
novelty search and BOGP,
as well as random search and BOGP.
Concerning the \textit{routing cost}, a statistically significant difference exists 
for 500 cars between
novelty search and BOGP,
novelty search and NSGA-II,
random search and NSGA-II,
as well as random search and novelty search;
for 700 cars only between
random search and BOGP,
while there is no statistically significant difference between any two techniques for 800 cars.

In general, we observe that by increasing cars, the average \textit{trip overhead} and \textit{routing cost} of the pareto-optimal configurations across all optimization techniques increase as well. This matches our expectation that with increasing traffic, it is more difficult to optimize the system in absolute terms.

Besides considering each objective individually, we want to investigate how \textit{both objectives together} are satisfied by solutions found by the different optimization techniques. Instead of defining a utility function over the \textit{trip overhead} and \textit{routing cost}, which may introduce some bias toward one objective, we use the well-known quality indicator \textit{hypervolume}~\cite{Wang:2016}.\footnote{We use variant 3 of the hypervolume algorithm by Fonseca et al.~\cite{Fonseca:2006}. Implementation: \url{https://ls11-www.cs.tu-dortmund.de/rudolph/hypervolume/}.}
In general, the hypervolume measures the volume in the objective space that is dominated by a pareto front. Thus, a higher hypervolume indicates a pareto front of better quality.

Thus, we computed the hypervolume for each pareto front of the 30 replicates for each technique and situation. The resulting data is plotted in Figures~\ref{sf:hv-500}-\subref{sf:hv-800} and the average and mean hypervolumes are listed in Table~\ref{tab:results}. Since the average and median values do not differ much, we consider the average hypervolume in the following.
For the situations with 500 and 700 cars, the pareto-front found by NSGA-II achieves the highest average hypervolume (173.09 and 114.56), followed by random search (172.41 and 113.77), BOGP (171.11 and 111.62), and novelty search (170.44 and 112.91). 
In contrast, for 800 cars the highest average hypervolume is achieved by the pareto front obtained by random search (85.00), closely followed by novelty search (84.42) and NSGA-II (84.39), and finally BOGP (82.43).
Concerning the hypervolume, we notice a statistically significant difference
for 500 cars between
\mbox{NSGA-II} and BOGP,
NSGA-II and novelty search,
as well as novelty search and random search;
for 700 cars between
NSGA-II and BOGP,
as well as random search and BOGP;
and for 800 cars between
random search and BOGP.

\subsubsection{Convergence}
We evaluate the convergence of the different techniques by evaluating how the hypervolume evolves
by plotting the achieved hypervolume over the 100 fitness evaluations for each situation as shown in Figures~\ref{sf:hve-500}-\subref{sf:hve-800}.
While for 500 cars, there is no distinct difference between the techniques, we observe for 700 and 800 cars that BOGP achieves slightly better results during the first 10 fitness evaluations than the other techniques. However, later on BOGP converges faster than the other techniques.  Thus, BOGP is a promising technique to find good configurations quicker, which supports a faster adaptation cycle while the other techniques may continue their search to find better configurations used for an adaptation later on in time. 
For instance, one run of NSGA-II with a budget of 1000 fitness evaluations (ten times the budget we considered so far) achieved a hypervolume of 186.14 for 500 cars. This run illustrates that better solutions can be obtained by NSGA-II with a larger budget. However, a budget of 1000 evaluations corresponds to an optimization horizon of 600 minutes (when evaluating 10 configurations in parallel), which prevents a timely adaptation in a traffic~system.

\subsubsection{Overhead}

We now discuss the overhead of each optimizer. As such, Figures~%
\ref{sf:mem} and~\ref{sf:proc} present our performance metrics for 500, 700, and 800 car counts, and Table~\ref{tab:perf-results} summarizes those results.  Note that, for each plot, the optimizers are presented in the order of 
\mbox{NSGA-II}, novelty search, and random search, respectively.  We examine the
peak memory overhead (kb) and peak processor usage (\%) required to execute RTX.

\newcommand{\perfwidth}{2.75in}
\begin{figure}[tb!]%
\centering
\vspace{-1mm}
  \subfigure[Peak memory usage (kb).]{%
  \label{sf:mem}%
  \includegraphics[width=\perfwidth]{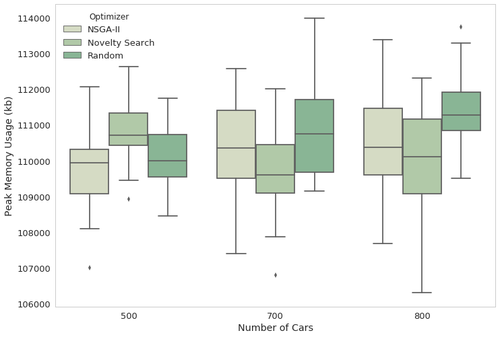}}%
  \qquad\vspace{-1mm}
  \subfigure[Peak processor usage (\%)]{%
  \label{sf:proc}%
  \includegraphics[width=\perfwidth]{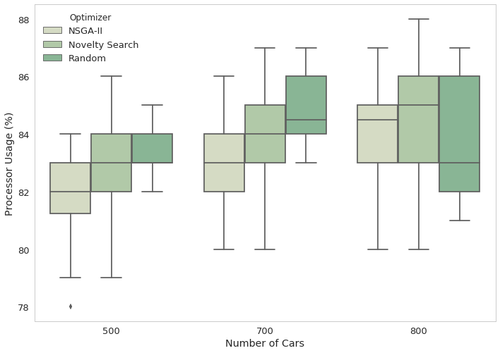}}%

  \vspace{-0.1in}

  \caption{Performance metrics for all evaluations.}
  \label{f:perf}
  \vspace{-1em}
\end{figure}

\begin{table}[t]
\caption{Memory Overhead and Processor Usage for 500, 700, 800 cars.}
\vspace{-2mm}
\resizebox{\columnwidth}{!}{%
\begin{tabular}{llrrrrrr}
& \textbf{Technique} & \multicolumn{2}{c}{\textbf{Memory Overhead (kb)}} & \multicolumn{2}{c}{\textbf{Processor Usage (\%)}} \\
& & Average & Median & Average & Median \\ \hline
\parbox[t]{2mm}{\multirow{3}{*}{\rotatebox[origin=c]{90}{\textbf{500}}}} 
& NSGA-II         & \cellcolor{gray!25}109687.2000 & \cellcolor{gray!25}109944.0000 & \cellcolor{gray!25}82.0333 & \cellcolor{gray!25}82.0000\\
& Novelty Search  & 110802.1333 & 110710.0000 & 82.6667 & 83.0000\\
& Random          & 110135.7333 & 109994.0000 & 83.4000 & 83.0000\\ \hline
\parbox[t]{2mm}{\multirow{3}{*}{\rotatebox[origin=c]{90}{\textbf{700}}}}  
& NSGA-II         & 110308.6667 & 110352.0000 & \cellcolor{gray!25}83.0333 & \cellcolor{gray!25}83.0000\\
& Novelty Search  & \cellcolor{gray!25}109671.6000 & \cellcolor{gray!25}109594.0000 & 84.0333 & 84.0000\\
& Random          & 110873.4667 & 110750.0000 & 84.8667 & 84.5000\\ \hline
\parbox[t]{2mm}{\multirow{3}{*}{\rotatebox[origin=c]{90}{\textbf{800}}}}  
& NSGA-II         & 110543.6000 & 110360.0000 & \cellcolor{gray!25}84.1667 & 84.5000\\
& Novelty Search  & \cellcolor{gray!25}109932.2667 & \cellcolor{gray!25}110096.0000 & 84.5000 & 85.0000\\
& Random          & 111425.7333 & 111274.0000 & \cellcolor{gray!25}84.1667 & \cellcolor{gray!25}83.0000\\ \hline
\end{tabular}%
}
\label{tab:perf-results}
\vspace{-2em}
\end{table}

For the presented metrics, a general trend of increasing overhead is seen as the number of cars increases, with random search tending to require the most resources.   
For both memory overhead and processor usage, we see no statistically significant difference between random, NSGA-II, and novelty search at any of the situations. %
As such, these results suggest that each of our implemented optimizers, including the baseline, require a similar amount of memory and CPU overhead.

\subsubsection{Discussion}
Given the results from evaluating multiple optimization techniques, we see that these techniques struggle with optimizing the \textit{trip overhead}. Our interpretation of these results is that:
(1)~We investigated the valuations of the input parameters for the pareto-optimal configurations across all techniques. We found that these valuations are spread in the search space so that we can assume that there are many local minima.
(2)~The trip overhead is influenced by all of the seven input parameters, which results in a large search space.
(3)~The trip overhead is rather noisy (it has high variance). 
These three aspects make it difficult to optimize the trip overhead.
In contrast, the \textit{routing cost} is easier to optimize than the \textit{trip overhead} as it is only affected by one input parameter (\textit{re-routing frequency}). Therefore, a technique might identify and follow a gradient based on the relationship that a higher routing frequency leads to lower routing cost.

Considering the goal of selecting one optimization technique, it depends on which criterion the selection is based on. 
If it is based on the solution quality, NSGA-II performs best~--~even though slightly~--~in two situations (500 and 700 cars) and only slightly worse than the best technique in the remaining situation (800 cars).
Nevertheless, random search performs surprisingly well in comparison to the other, more intelligent techniques. A reason for this might be the many local minima that exist for the \textit{trip overhead} (\textit{cf.}~previous paragraph) so that a random search may easily catch such a minimum with 100 trials of random configurations.
Similar observations, that random search performs well in cases of parameter optimization, have been made~\cite{Seymour:2008,Bergstra:2012} and witness that random search can be an effective technique for optimizing high-dimensional, black-box systems.
Considering the convergence of the different techniques, BOGP should be selected since it finds good configurations quicker than the other techniques. However, it shows smaller improvement in the solution quality in longer runs.
Finally, no selection can be done based on the overhead of the different optimization techniques, since their overhead in terms of memory and processor usage is comparable.

\subsection{Threats to Validity} 

We have identified the following threats to validity of the evaluation results.
First, we have used only a single context parameter to show the feasibility of runtime clustering for situation detection in CrowdNav.
We further rely on the well-known $k$-means clustering algorithm for situation detection, using the Silhouette method for determining the best value of $k$. 
Other methods exist to learn the optimal number of clusters, such as XMeans~\cite{Pelleg2000}, that may lead to different results.
Second, when optimizing for a situation, we set the number of vehicles to the largest number in the corresponding cluster, assuming that this is representative of other vehicle numbers in the cluster.
Third, we have used the vanilla version of the three optimization strategies we selected.  
Thus, we did not tune the meta-parameters (e.g., number of generations, crossover rate, etc.) to tailor each technique specifically to CrowdNav.
Fourth, this study focused on CrowdNav as a representative of the class of systems corresponding to black-box, high-dimensional, and expensive problems. 
Therefore, although our approach of planning as optimization may generalize to other systems in this class, the evaluation results are obtained for a specific simulated system (CrowdNav) and cannot be generalized to other systems.

\section{Challenges}
\label{sec:implications}

We have presented a proof of concept of the planning as optimization approach, together with an empirical study of different optimization techniques applied in a complex system that corresponds to a black-box, high-dimensional, and computationally expensive optimization problem.
Our evaluation results indicate that none of the compared techniques is superior in optimizing CrowdNav in terms of solution quality, convergence, and overhead. 
Moreover, the results indicate that different techniques perform better in different situations of the running system with respect to different objectives. 
Thus, these insights motivate our vision of \textit{self-learning continuous optimization}: to use multiple optimization techniques at runtime and switch between them according to the situation and objective of optimization, while always having an optimization process and a situation identification process running.
To realize the vision, our proposed approach must be extended by addressing the following challenges.

\textbf{Continuous clustering.}
While performing clustering at runtime based on system outputs to identify distinct situations, the number and range of situations may evolve in time. 
For instance, in the first 30 minutes of collecting output data three situations may be identified; this number may evolve to four after 60 minutes. 
These four situations may even have no overlap with the previous three. 
An approach for self-learning continuous optimization should be able to detect situations that do not change, or similar situations between consecutive learning phases for which optimal configurations can be reused. 
Moreover, it should effectively ``forget" old data to identify clusters that correspond to the latest environment.

\textbf{Seamless operation of optimizers.}
Our planning as optimization approach needs to be able to pause an optimization process when the current situation changes and continue it when the situation arises again. 
For instance, when the current situation $s_a$ changes to $s_b$ while optimizer $o_a$ is running, $o_a$ needs to store its status (e.g., the best solutions found so far) to the \texttt{Knowledge Base} (Figure~\ref{fig:approach}) to reuse it when $s_a$ appears again. 
Self-learning continuous optimization not only needs to be able to pause and resume the operation of optimizers, but also dynamically switch between optimizers at runtime.

\textbf{Automated comparison of optimizers.}
In our empirical study, we have compared three optimizers based on solution quality, convergence, and overhead. 
We presented all the results and drew conclusions which can be used for choosing one optimizer over another. 
In self-learning continuous optimization, such conclusions need to be taken by the system itself, which raises a number of challenges: 
How many iterations to perform per optimizer? 
How many samples to collect for the evaluation of a configuration? 
Which criteria to use in the comparison?
Consider also that different situations (e.g., accidents) may require a change in the choice of optimizers (e.g., select the fast and less effective optimizer).

\section{Conclusion}
\label{sec:conclusion}

In this paper, we presented the \textit{planning as optimization} approach that uses optimization strategies to discover optimal system configurations at runtime for each distinct situation that is dynamically identified at runtime. 
We instantiated our approach with well-known techniques such as the k-means clustering algorithm to identify distinct situations, and Bayesian optimization with Gaussian Processes (BOGP), NSGA-II, and novelty search for finding optimal configurations.
Our approach tackles complex, real-world systems such as CrowdNav that can be modeled as black-box, high-dimensional, and computationally expensive optimization problems. 

We show the feasibility of planning as optimization by dynamically identifying distinct situations via clustering and by identifying optimal configurations via optimization techniques. We further compare the solution quality, convergence, and overhead of three optimization techniques in an empirical study with CrowdNav.
The results show that no technique is \textit{significantly} superior for all three situations in terms of the solution quality. 
However, NSGA-II performs slightly better in terms of solution quality in two situations while BOGP converges faster in all three situations. 
With respect to CPU and memory overhead, no technique is significantly different.

Finally, we discussed our vision of self-learning continuous optimization and related open research challenges:
(i)~continuous clustering;
(ii)~seamless operation of optimizers; and
(iii)~automated comparison of optimizers.

\section*{Acknowledgment}

This research has been supported in part by NSF grant CNS-1657061, the Michigan Space Grant Consortium, the Comcast Innovation Fund, Google Cloud, and Oakland University.
This work is part of the ViM project funded by the Bavarian Ministry of Economic Affairs, Regional Development and Energy.
This work has been partly developed in the \textit{FLASH} project funded by the German Science Foundation~(DFG).

\bibliographystyle{IEEEtran}
\balance
\bibliography{SASO19_Optimization}

\end{document}